\documentclass{appolb}

\def\la{\mathrel{\mathchoice {\vcenter{\offinterlineskip\halign{\hfil
$\displaystyle##$\hfil\cr<\cr\sim\cr}}}
{\vcenter{\offinterlineskip\halign{\hfil$\textstyle##$\hfil\cr<\cr\sim\cr}}}
{\vcenter{\offinterlineskip\halign{\hfil$\scriptstyle##$\hfil\cr<\cr\sim\cr}}}
{\vcenter{\offinterlineskip\halign{\hfil$\scriptscriptstyle##$\hfil\cr<\cr
\sim\cr}}}}}

\def\ga{\mathrel{\mathchoice {\vcenter{\offinterlineskip\halign{\hfil
$\displaystyle##$\hfil\cr>\cr\sim\cr}}}
{\vcenter{\offinterlineskip\halign{\hfil$\textstyle##$\hfil\cr>\cr\sim\cr}}}
{\vcenter{\offinterlineskip\halign{\hfil$\scriptstyle##$\hfil\cr>\cr\sim\cr}}}
{\vcenter{\offinterlineskip\halign{\hfil$\scriptscriptstyle##$\hfil\cr>\cr
\sim\cr}}}}}

\usepackage{epsfig}
\begin{document}
%\date{\today}
\pagestyle{plain}
%% uncomment the following line to get equations numbered by (sec.num)
%\eqsec
\newcount\eLiNe\eLiNe=\inputlineno\advance\eLiNe by -1

\title{Cosmic Rays in the `Knee'-Region\\
-- Recent Results from KASCADE --}
%\thanks{}%

\author{
K.-H.~Kampert$^{1}$\footnote{\tt email: 
kampert@uni-wuppertal.de},\\ 
T.~Antoni$^{2}$, 
W.D.~Apel$^{3}$, 
F.~Badea$^{3}$, 
K.~Bekk$^{3}$ , 
A.~Bercuci$^{3}$, 
M.~Bertaina$^{4}$, 
H.~Bl\"umer$^{3,2}$, 
H.~Bozdog$^{3}$, 
I.M.~Brancus$^{5}$, 
M.~Br\"uggemann$^{6}$, 
P.~Buchholz$^{6}$, 
C.~B\"uttner$^{2}$, 
A.~Chiavassa$^{4}$, 
A.~Chilingarian$^{7}$, 
K.~Daumiller$^{2}$, 
P.~Doll$^{3}$, 
R.~Engel$^{3}$, 
J.~Engler$^{3}$, 
F.~Fessler$^{3}$ , 
P.L.~Ghia$^{8}$, 
H.J.~Gils$^{3}$, 
R.~Glasstetter$^{1}$, 
A.~Haungs$^{3}$, 
D.~Heck$^{3}$, 
J.R.~H\"orandel$^{2}$ , 
H.O.~Klages$^{3}$, 
Y.~Kolotaev$^{6}$, 
G.~Maier$^{3}$, 
H.J.~Mathes$^{3}$ , 
H.J.~Mayer$^{3}$ , 
J.~Milke$^{3}$ , 
C.~Morello$^{8}$, 
M.~M\"uller$^{3}$ , 
G.~Navarra$^{4}$, 
R.~Obenland$^{3}$, 
J.~Oehlschl\"ager$^{3}$, 
S.~Ostapchenko$^{2}$, 
M.~Petcu$^{5}$ , 
S.~Plewnia$^{3}$, 
H.~Rebel$^{3}$, 
A.~Risse$^{9}$, 
M.~Risse$^{3}$, 
M.~Roth$^{2}$, 
G.~Schatz$^{3}$, 
H.~Schieler$^{3}$, 
J.~Scholz$^{3}$,
M.~St\"umpert$^{2}$, 
T.~Thouw$^{3}$, 
G.C.~Trinchero$^{8}$, 
H.~Ulrich$^{3}$, 
S.~Valchierotti$^{4}$, 
J.~van~Buren$^{3}$, 
A.~Vardanyan$^{7}$, 
W.~Walkowiak$^{6}$, 
A.~Weindl$^{3}$, 
J.~Wochele$^{3}$, 
J.~Zabierowski$^{9}$, 
S.~Zagromski$^{3}$
% 
%}              % Do not remove
%
\address{%
$^{1}$ Fachbereich C - Physik, Universit\"at Wuppertal, 42097
Wuppertal, Germany\\
$^{2}$ Institut f\"ur Exp.\ Kernphysik,
Universit\"at Karlsruhe, 76021 Karlsruhe, Germany\\
$^{3}$ Inst.\ f\"ur Kernphysik, Forschungszentrum Karlsruhe,
76021~Karlsruhe, Germany\\
$^{4}$ Dipartimento di Fisica Generale dell'Universit{\`a},
10125 Torino, Italy\\
$^{5}$ National Inst.\ of Physics and Nuclear Engineering,
7690~Bucharest, Romania\\
$^{6}$ Fachbereich Physik, Universit\"at Siegen, 57072 Siegen, 
Germany\\
$^{7}$ Cosmic Ray Division, Yerevan Physics Institute, Yerevan~36, 
Armenia\\
$^{8}$ Istituto di Fisica dello Spazio Interplanetario, CNR, 
10133 Torino, Italy\\
$^{9}$ Soltan Institute for Nuclear Studies, 90950~Lodz, 
Poland\\
}}

\maketitle

\begin{abstract}
Recent results from the KASCADE experiment on measurements of
cosmic rays in the energy range of the knee are presented.
Emphasis is placed on energy spectra of individual mass groups as
obtained from sophisticated unfolding procedures applied to the
reconstructed electron and truncated muon numbers of EAS. The
data show a knee-like structure in the energy spectra of light
primaries (p, He, C) and an increasing dominance of heavy ones
($A>20$) towards higher energies.  This basic result is robust
against uncertainties of the applied interaction models QGSJET
and SIBYLL. Slight differences observed between experimental data
and EAS simulations provide important clues for improvements of
the interaction models.  The data are complemented by new limits
on global anisotropies in the arrival directions of CRs and by
upper limits on point sources.  Astrophysical implications for
discriminating models of maximum acceleration energy vs galactic
diffusion/drift models of the knee are discussed based on this
data.  To improve the reconstruction quality and statistics
around $10^{17}$\,eV, KASCADE has recently been extended by a
factor 10 in area.  The status and expected performance of the
new experiment KASCADE-Grande is presented.
\end{abstract}

\section{Introduction}
\begin{figure}[t]
\centerline{\epsfxsize=11cm\epsfbox{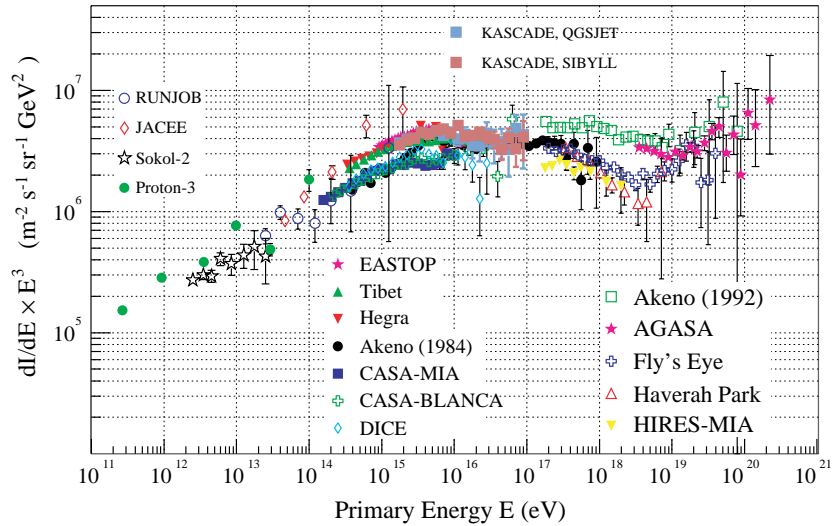}}
\caption[]{Compilation of the all-particle cosmic ray spectrum 
showing the knee, the suggested second knee, and the ankle of the 
CR spectrum (compiled by H. Ulrich).
\label{fig:all-particle-compilation}}
\end{figure}

A puzzling and most prominent feature of the cosmic ray (CR)
spectrum is the so-called knee, where the spectral index of the
all-particle power-law spectrum changes from approximately $-2.7$
to $-3.1$.  Several models have been proposed in order to explain
this feature shown in Fig.\,\ref{fig:all-particle-compilation},
but none of them has managed to become broadly accepted.  Some
models focus on a possible change in the acceleration mechanism
at the knee~\cite{Lagage83,Biermann93,Drury94}, e.g.\ due to the
limiting energy defined by the size and magnetic field strength
of the acceleration region ($E_{\rm max} \la Z \times (B \times
L))$.  Other ones discuss an increased leakage of CRs from the
Galaxy due to a change in the confinement efficiency by galactic
magnetic fields~\cite{Wdowczyk84,Candia02a,Roulet03}.  Again,
this results in a rigidity scaling of the knee according to the
maximum confinement energy.  Finally, a third group of models
attributes the effect of the knee to CR interactions at their
sources, during their propagation in the Galaxy, or in the upper
atmosphere.  Such scenarios include nuclear photodisintegration
processes by UV-photons at the sources~\cite{Candia02b},
interactions of CRs in dense fields of massive relic
neutrinos~\cite{Wigmans03}, production of gravitons in
high-energy pp collisions~\cite{Kazanas01}, etc.  A recent review
about this topic can be found e.g.\ in
Refs.~\cite{Kampert01a,Haungs03}.

To distinguish between these models and allowing to answer the
long pressing question about the origin of cosmic rays and about
the knee in their spectrum, high quality and high statistics data
are required over an energy interval ranging from at least 0.5 to
500 PeV. It appears worthwhile to mention that solving the old
problem about the origin of CRs in the PeV region is a
prerequisite also for an understanding of the highest energies in
the GZK-region.  Due to the low flux involved, only extensive air
shower (EAS) experiments are able to provide such data.  In EAS
experiments, primary CRs are only indirectly observed via their
secondaries generated in the atmosphere.  The most important
experimental observables at ground are then the electromagnetic
(electrons and photons), muonic, and hadronic components.  In
addition or alternatively, some experiments also detect photons
originating from Cherenkov and/or fluorescence radiation of
charged particles in the atmosphere.  For a brief review about
EAS observables and their experimental techniques the reader is
referred to Refs.\,\cite{Kampert01b,Swordy02,Haungs03}.

Unfortunately, progress on interpreting EAS data has been 
modest mostly because of two reasons: Firstly, the EAS development
is driven both by the poorly known high-energy hadronic
interactions and their particle production in the very forward
kinematical region as well as by uncertainties in the low energy
interaction models influencing mostly the lateral particle
density distribution functions \cite{Drescher04}.  Secondly, due
to the stochastic nature of particle interactions, most
importantly the height of the very first interaction in the
atmosphere, EAS are subject to large fluctuations in particle
numbers at ground.  To make things even more complicated, the
amount of fluctuations depends, amongst others, sensitively on
the primary CR energy and mass~\cite{Kampert01b}.  Here, it is
very important to realize that EAS fluctuations are not to be
mistaken as random Gaussian errors associated with the statistics
in the number of particles observed at ground.  The latter one
can be improved by the sampling area of an EAS experiment, while
the former one is intrinsic to the EAS itself, carrying - for a
sample of events - important information about the nature of the
primary particle.  Clearly, both kinds of fluctuations have to be
accounted for in the data analysis of steeply falling energy
spectra in order to not misinterpret the observations.

\section{Results from the KASCADE Experiment}
KASCADE (\underline{Ka}rlsruhe \underline{S}hower
\underline{C}ore and \underline{A}rray \underline{De}tector) is a
sophisticated EAS experiment for detailed investigations of
primary CRs in the energy range of the knee.  For reconstructing
the CR energy and mass and for investigating high-energy hadronic
interactions, KASCADE follows the concept of a multi-detector
set-up providing as much complementary information as possible as
well as redundancy for consistency tests.  Most relevant for the
results presented in this paper is the scintillator array
comprising 252 detector stations of electron and muon counters
arranged on a grid of $200 \times 200$ m$^{2}$.  In total, it
provides about 500 m$^2$ of $e/\gamma$- and 620 m$^{2}$ of
$\mu$-detector coverage.  The detection thresholds for vertical
incidence are $E_{e} > 5$ MeV and $E_{\mu} > 230$ MeV. More
details about the $e/\gamma$- and $\mu$-detectors and all other
other detector components can be found in
Ref.\,\cite{KASCADE-NIM}.

\subsection{Chemical Composition and Energy Spectra}

\begin{figure}[t]
\vspace*{-2mm}
\centerline{\epsfxsize=9cm\epsfbox{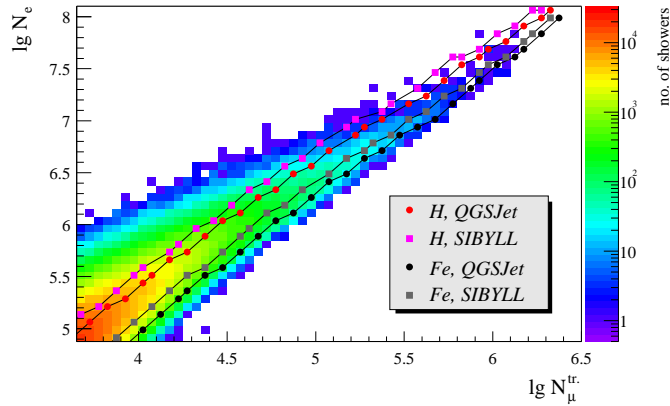}}
\vspace*{-2mm} \caption[]{Two dimensional electron ($N_{e}$) and
truncated muon number ($N_{\mu}^{\rm tr}$) spectrum measured by
the KASCADE array.  Lines display the most probable values
expected for proton and iron primaries according to CORSIKA
simulations employing two different hadronic interaction
models~\cite{Ulrich03}.
\label{fig:2d}}
\end{figure}

The traditional and perhaps most sensitive technique to infer the
CR composition from EAS data is based on measurements of the
electron ($N_e$) and muon numbers ($N_\mu$) at ground.  It is
well known~\cite{Kampert01b} that for given energy, primary
Fe-nuclei result in more muons and fewer electrons at ground as
compared to proton primaries.  Specifically, in the energy range
and at the atmospheric depth of KASCADE, a Fe-primary yields
about 30\,\% more muons and almost a factor of two fewer
electrons as compared to a proton primary.  The basic
quantitative procedure of KASCADE for obtaining the energy and
mass of the cosmic rays is a technique of unfolding the observed
two-dimensional electron-vs-muon number spectrum of
Fig.\,\ref{fig:2d} into the energy spectra of primary mass
groups~\cite{Ulrich03}.  The problem can be considered a
system of coupled Fredholm integral equations of the form
\begin{equation}
\frac{dJ}{d\,\lg N_e \;\; d\,\lg N_\mu^{\rm tr}} = 
  \sum_A \int\limits_{-\infty}^{+\infty} \frac{d\,J_A}{d\,\lg E} 
  \cdot 
  p_A(\lg N_e\, , \,\lg N_\mu^{\rm tr}\, \mid \, \lg E)
  \cdot 
  d\, \lg E
\end{equation}  
where the probability $p_A$  \\
{\small   $p_A(\lg N_e , \lg N_\mu^{\rm tr}\, \mid \, \lg E) =
    \int\limits_{-\infty}^{+\infty} k_A(\lg N_e^t , \lg 
    N_\mu^{\rm tr,t})
  d\, \lg N_e^t\,\, d\,\lg N_\mu^{\rm tr,t} $ } \\
is another integral equation with the kernel function $k_A = r_A
\cdot \epsilon_A \cdot s_A$ factorizing into three parts.  Here,
$r_A$ describes the shower fluctuations, i.e.\ the 2-dim
distribution of electron and truncated muon number for fixed
primary energy and mass, $\epsilon_A$ describes the trigger
efficiency of the experiment, and $s_A$ describes the
reconstruction probabilities, i.e.\ the distribution of $N_e$ and
$N_\mu^{\rm tr}$ that is reconstructed for given {\em true}
numbers $N_e^t$, $N_\mu^{\rm tr,t}$ of electron and truncated
muon numbers.  The probabilities $p_A$ are obtained by
parameterizations of EAS Monte Carlo simulations for fixed
energies using a moderate thinning procedure as well as smaller
samples of fully simulated showers for the input of the detector
simulations.  Because of the shower fluctuations mentioned above,
unfolding of all 26 energy spectra ranging from protons to
Fe-nuclei is clearly impossible.  Therefore, 5 elements (p, He,
C, Si, Fe) were chosen as representatives for the entire
distribution.  More mass groups do not improve the
$\chi^2$-uncertainties of the unfolding but may result in mutual
systematic biases of the reconstructed spectra \cite{Ulrich03}.

The unfolding procedure is tested by using random initial spectra
generated by Monte Carlo simulations.  It has been
shown~\cite{Ulrich03} that knee positions and slopes of the
initial spectra are well reproduced and that the discrimination
between the five primary mass groups is sufficient.  For
scrutinizing the unfolding procedure, different mathematical ways
of unfolding (Gold-algorithm, Bayes analyses, principle of
maximum entropy, etc.)  have been compared and the results are
consistent~\cite{Ulrich03}.  For generating the kernel functions
a large number of EAS has been
simulated~\cite{Ulrich03,roth-ulrich} employing
CORSIKA~\cite{cors} with the hadronic interaction models QGSJET
(version 2001)~\cite{qgs} and SIBYLL 2.1~\cite{sib}.

\begin{figure}[t]
\centerline{\epsfxsize=\textwidth\epsfbox{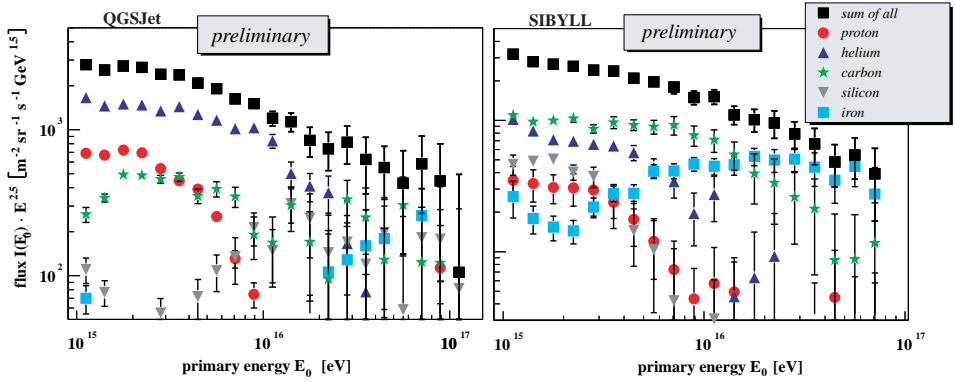}}
\caption[]{Results of the unfolding procedure using QGSJET (left) 
and SIBYLL (right) as hadronic interaction model~\cite{Ulrich03}.
\label{fig:spectra}}
\end{figure}

The result of the unfolding is presented in
Fig.\,\ref{fig:spectra} for each of the two interaction models.
Clearly, there are common features but also differences in the
energy distributions obtained with the two interaction models.
The all-particle spectra coincide very nicely and in both cases
the knee is caused by the decreasing flux of the light primaries,
corroborating results of an independent analysis of
Ref.\,\cite{muon-density}.  Tests using different data sets,
different unfolding methods, etc.\ show the same
behavior~\cite{roth-ulrich}.  As the most striking difference,
SIBYLL suggests a more prominent contribution of heavy primaries
at high energies.  This difference results from the different
$N_{e}$-$N_{\mu}^{\rm tr}$ correlation shown in
Fig.\,\ref{fig:2d}, i.e.\ SIBYLL predicts higher electron and
lower muon numbers for given primaries as compared to QGSJET.

\begin{figure}[t]
%\centerline{\epsfxsize=\textwidth\epsfbox{chi2planeqgs-sibyll.eps}}
\centerline{\epsfxsize=\textwidth\epsfbox{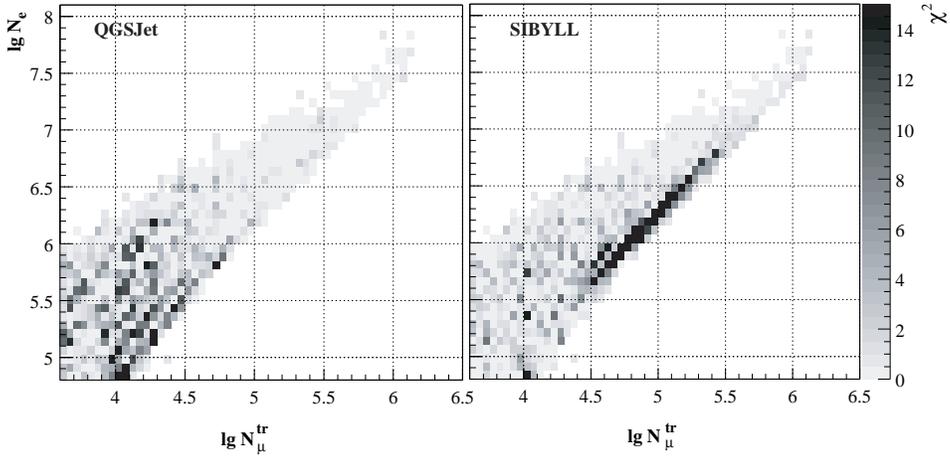}}
\caption[]{$\chi^{2}$ deviation between the forward folded 
reconstructed and measured  
($N_{e}$-$N_{\mu}^{\rm tr}$)-data cells for the QGSJET (left) 
and SIBYLL (right) hadronic interaction models~\cite{Ulrich03}.
\label{fig:chi2}}
\end{figure}

Is there a way to judge which of the two models is better suited
for describing the data?  This is done most easily by comparing
the residuals of the unfolded two-dimensional $N_{e}$ vs
$N_{\mu}^{\rm tr}$ distributions with the actual data used as
input to the unfolding (Fig.\,\ref{fig:2d}).  The result of such
an analysis is presented in Fig.\,\ref{fig:chi2} in terms of
$\chi^2$.  The deviations seen reveal some deficiencies of QGSJET
at low electron and muon numbers and they nicely demonstrate that
SIBYLL encounters problems in describing the high-$N_{e}$ -
low-$N_{\mu}^{\rm tr}$ tail of the experimental data at 10 PeV
and above~\cite{Ulrich03}.  If not being prepared to accept an
additional significant contribution of superheavy primaries ($A >
60$) required in case of SIBYLL simulations to fill the gap at
high muon numbers, the results point to a muon deficit (a/o
electron abundance) in this model.  Definitely, this problem
needs further attention and will be very important also for
composition studies at higher energies~\cite{Watson03}.

With this caveats kept in mind, the KASCADE data favor an
astrophysical interpretation of the knee and are in agreement
with a constant rigidity of the knee position for the different
primaries.  Similar results were very recently obtained from
combined EAS-TOP / MACRO measurements~\cite{Aglietta03}, though
for two mass groups only, and were again confirmed for three mass
groups from EAS-TOP electron and muon
measurements~\cite{Navarra04}.  Within the given error bars, the
mean logarithmic masses of both experiments agree well with one
another.

\subsection{Search for Anisotropies and Point Sources}

Additional information about the CR origin and their propagation
in the galactic environment can be obtained from global
anisotropies in their arrival directions.  Model calculations
show that diffusion of CRs in the galactic magnetic field can
result in anisotropies on a scale of $10^{-4}$ to $10^{-2}$
depending on the particle energy and on the strength and
structure of the galactic magnetic field \cite{Candia02a}.  Since
the diffusion scales again with the rigidity, a factor of 5-10
larger anisotropies are expected for protons as compared to iron
primaries.  This rigidity dependent diffusion is one of the
possible explanations of the knee.

\begin{figure}[t]
\centerline{\epsfxsize=\textwidth\epsfbox{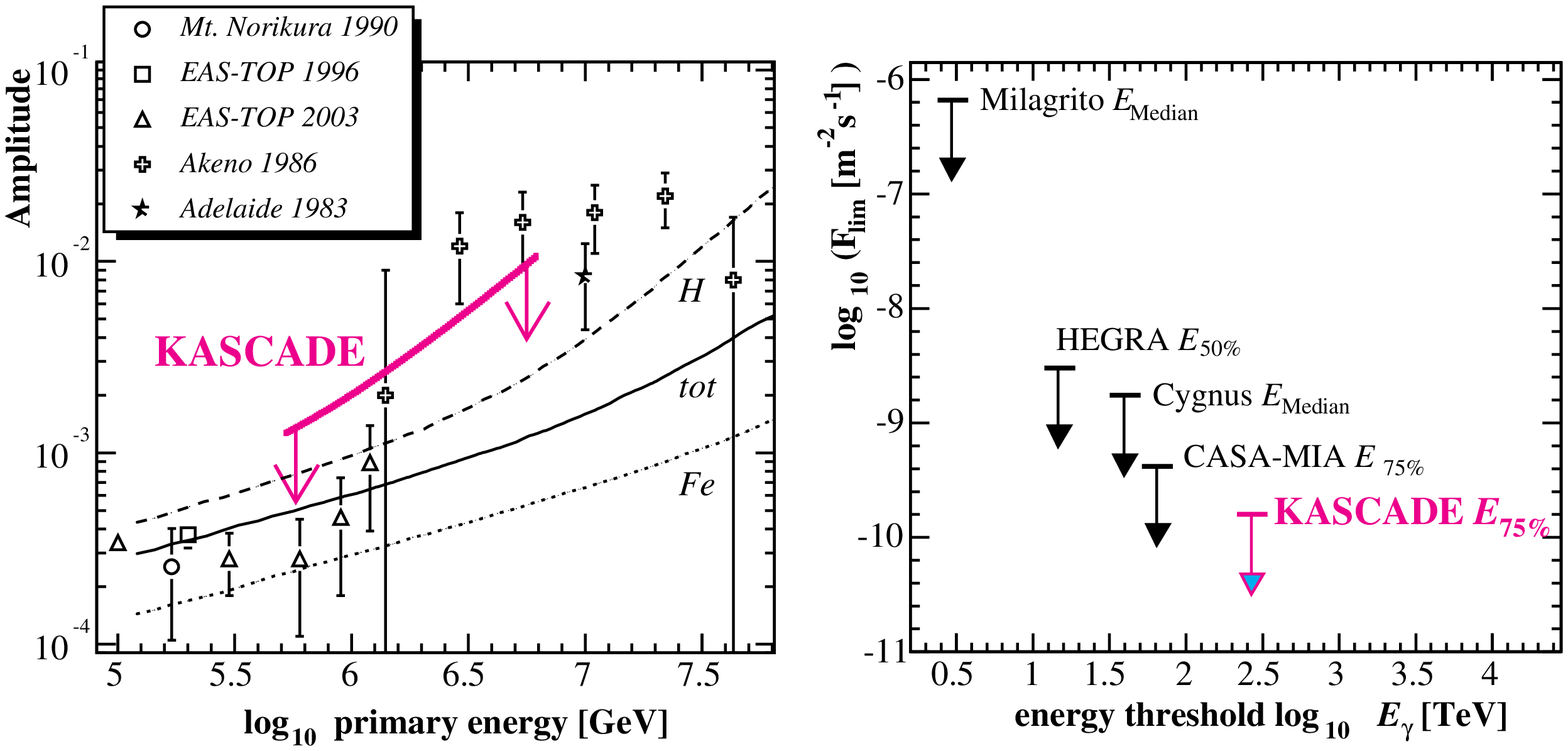}}
\caption[]{Left: KASCADE upper limits (95\,\%) of Rayleigh
amplitudes $A$ vs primary energy (bold line) compared to results
from other experiments and to expectations from galactic CR
diffusion~\cite{global-anisotropies}.\\
Right: 90\,\% upper limit for a point source moving through the
zenith in comparison with results from other
experiments~\cite{point-sources}.  Note the different definitions
of the energy thresholds.
\label{fig:anisotropy}}
\end{figure}

Due to the small anisotropy expected a large data sample and
careful data selection is necessary.  About $10^{8}$ EAS events
in the energy range from 0.7 to 6 PeV were selected and studied
in terms of Rayleigh amplitudes $A$ and phases $\Phi$ of the first
harmonic.  Neither for the full set of data nor for electron-rich
and -poor EAS significant Rayleigh amplitudes were found. The 
upper limit on the large scale anisotropy is depicted in 
Fig.\,\ref{fig:anisotropy} \cite{global-anisotropies} and is in 
line with results reported from other experiments. We shall come 
back to this result in the next section. 

Even though the location of CR sources should be obscured due to
the deflection of charged particles in the magnetic field of our
galaxy, there is interest to perform point source searches.  For
example, neutrons are not deflected and can reach the Earth if
their energy and hence decay length is comparable with the
distance of the source.  A decay length of 1 kpc corresponds to a
neutron energy of about $10^{17}$ eV. Also, by applying
appropriate cuts to electron and muon numbers from EAS, searches
for $\gamma$-ray point sources can be performed in the PeV range.

Such a study has been performed based on 47 Mio EAS with primary
energies above $\sim 300$ TeV. A certain region in the sky is
then analyzed by comparing the number of events from the assumed
direction with an expected number of background events.  For the
latter, the so-called time-shuffling method has been used.  As a
result, again no significant excess is found in the region of the
galactic plane or for selected point source candidates.  Assuming
equal power laws in the energy spectra of background and source
events, upper flux limits can be calculated for given energy
thresholds.  For a steady point source that transits the zenith,
we obtain an upper flux limit of $3 \cdot 10^{-10}$
m$^{-2}$s$^{-1}$ (see Fig.\,\ref{fig:anisotropy} r.h.s.)
\cite{point-sources}. This is roughly 1-2 orders of magnitude 
larger than the Crab flux extrapolated to this energy.

Very recently, Chilingarian \etal reported the detection of a
source of high-energy CRs in the Monogem ring
\cite{chili-source}.  Changing slightly our cuts in zenith angle
to widen the declination range thereby covering the position of
the source candidate, we find 742 events within an opening angle
of $0.5^{\circ}$ around the suggested location with an expected
number of 716 background events yielding an upper flux limit of
$3 \cdot 10^{-10}$ m$^{-2}$s$^{-1}$.  Similar values are found
when searching for an excess from the location of the pulsar PSR
B0656+14 located near the centre of the Monogem SNR
\cite{Thorsett03}.

\subsection{Implications for understanding the CR origin}

The new high quality data presented in the previous sections have
revitalized the interest to understand both the origin of CRs in
the knee region and the phenomenon of the knee structure itself.

\begin{figure}[b]
\centerline{\epsfxsize=\textwidth\epsfbox{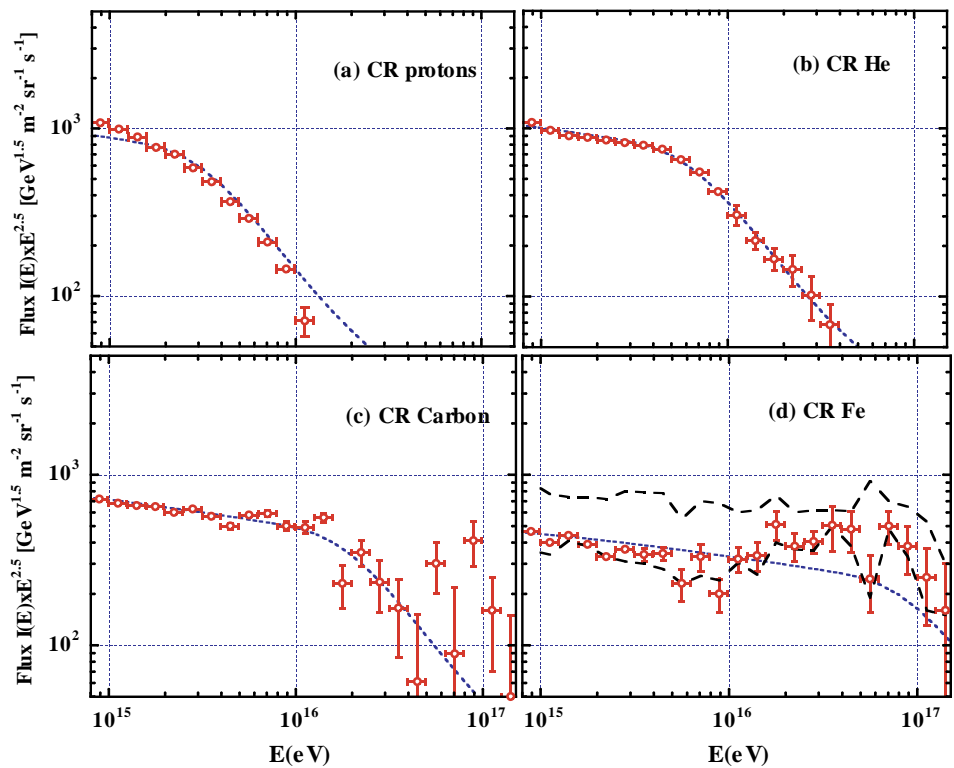}}
\caption[]{Fit of the GRB-model of Wick \etal \cite{Wick04} to
the preliminary KASCADE data presented at ICRC 2001
\cite{Kampert01c,Ulrich01}.  In the model, a GRB occurred
210\,000 years ago at a distance of 500 pc and injected
$10^{52}$~ergs into CRs. The CRs isotropically diffuse with an
energy-dependent mean-free-path in a MHD turbulence field.
\label{fig:wick-spectra}}
\end{figure}

This is because discriminating models of maximum acceleration
energy from galactic diffusion/drift models of the knee or from
particle physics interpretations require detailed inspection of
knee structures seen in {\em individual} mass groups combined
with precise measurements of CR anisotropies.  Targets of
particular interest are the individual energies of the spectral
break, the power-law indices below and above the knee, and the
smoothness of the turn-over regions.  Even though, these goals
are not yet achieved totally, important steps have been made to
it.  Previous investigations were limited to inclusive CR
all-particle spectra and to global changes of the mean
logarithmic mass, $\ln A$, with primary energy.  Unfortunately,
such measurements appear to be too insensitive for a convincing
discrimination of models.  Furthermore, analyses were mostly
restricted to investigating {\em mean} values of distributions to
be compared to EAS simulations.  This implies that deficiencies
of hadronic interaction models easily remain unrecognized.

A very good example demonstrating the discrimination power of the
new data presented here and showing the amount of information
contained in it is given by a recent study of Wick \etal
\cite{Wick04}.  Based on the earlier suggested connection between
Gamma-Ray Bursts (GRBs) and ultrahigh-energy CRs
\cite{Vietri95,Waxman95} they propose a model for the origin of
CRs from $\sim 10^{14}$~eV/nucleon up to the highest energies
($\ga 10^{20}$~eV).  In that model, GRBs are assumed to inject CR
protons and ions into the interstellar medium of star-forming
galaxies - including the Milky Way - with a power-law spectrum
extending to a maximum energy $\sim 10^{20}$~eV. High-energy CRs
injected in the Milky Way diffuse and escape from our Galaxy.
Ultra high-energy CRs with energies $\ga 10^{17}$ to
$10^{18}$\,eV that have Larmor radii comparable to the size scale
of the galactic halo escape directly from the Milky Way and
propagate almost rectilinearly through extragalactic space.  By
the same token, UHECRs produced from other galaxies can enter the
Milky Way to be detected.  UHECRs formed in GRBs throughout the
universe then travel over cosmological distances and have their
spectrum modified by energy losses, so an observer in the Milky
Way will measure a superposition of UHECRs from extragalactic
GRBs and HECRs produced in our Galaxy.

Thereby, the CR spectrum near the knee is understood by CRs
trapped in the Galactic halo that were accelerated and injected
by an earlier Galactic GRB. Assuming magneto-hydrodynamical
turbulence superposed to the galactic magnetic field, a fit to
the preliminary KASCADE data, shown in Fig.\
\ref{fig:wick-spectra}, suggests a 500 pc distant GRB
that released $10^{52}$~ergs in CRs if the GRB took place about
210\,000 yrs ago.  Keeping in mind the still large uncertainties
of the data and some freedom of parameters in the model, there is
remarkable accordance observed.  In this model, the rigidity
dependence of knee position is caused by galactic modulation
effects.

Since in this model a single galactic GRB is responsible for most
of the CRs in the knee region, anisotropies in the arrival
directions are expected on different levels, depending on the
distance and age of the GRB. For example, the authors state that
if an anisotropy below $\sim 0.2$\,\% is confirmed, then a number
of implications follow.  Either we are located near a rather
recent GRB, which could be unlikely, or the CR energy release
from GRBs is larger than the one given above \cite{Wick04}.
Thus, improving the anisotropy limits of the previous section
would help to further pin down this model.

However, before starting to over-interpret the data, we should
emphasize again the influence of the interaction models to
extracted energy spectra.  More work is still needed to improve
the models and to arrive at smaller systematic uncertainties.  On
the other hand, the very valuable study of an interesting model
demonstrates the informational content reached by present data.

\section{Status of KASCADE-Grande}

\begin{figure}[b]
\centerline{\epsfxsize=\textwidth\epsfbox{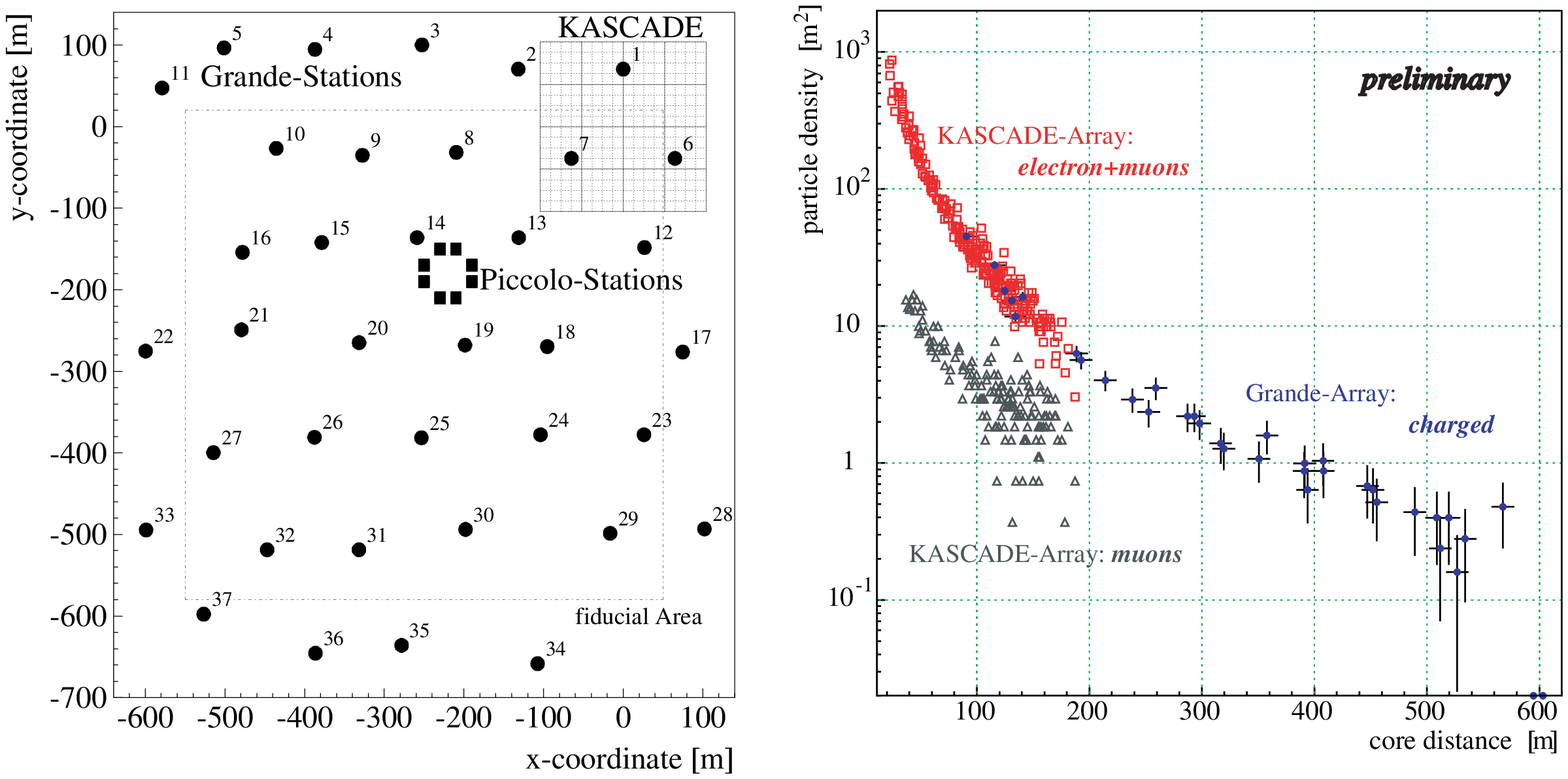}}
\caption[]{Left: Experimental set-up of KASCADE-Grande showing
the positions of its major components.  Right: Particle densities
observed with different detector components of KASCADE-Grande in
a single EAS.
\label{fig:KA-Gr-LDF}}
\end{figure}

The upper energy range of KASCADE is - besides event statistics -
mostly given by its sensitive area of $200 \times 200$ m$^{2}$
limiting the reconstruction quality of the highest energy events.
To improve the situation for EAS beyond $10^{17}$ eV, KASCADE has
recently been extended to KASCADE-Grande by an installation of
additional 45 detector stations (37 as {\sl Grande} array plus 8
as {\sl Piccolo} trigger array) over an area of $700 \times
700\,$m$^2$.  A sketch of the experimental set-up is depicted in
Fig.\,\ref{fig:KA-Gr-LDF} (lhs).  In the present configuration
KASCADE-Grande consists of $965\,$m$^2$ of scintillator area for
the electron component, of $1070\,$m$^2$ for measuring muons at
four different muon energy thresholds, and of $300\,$m$^2$ for
high-energy hadron detection.  Thus, KASCADE-Grande displays the
full capability of a multi-detector experiment with much better
muon sampling than any previous EAS experiment in this energy
range~\cite{kg}.  Data taking has started in July 2003 and an
example of lateral particle density distributions observed in a
single EAS is shown in Fig.\,\ref{fig:KA-Gr-LDF} (rhs).  The data
quality and detector performance is evident.  Not shown are muon
densities measured additionally with the KASCADE central detector
and with the muon tracking detector.  Sensitivity to the primary
mass is again given by the electron-muon density measurements as
well as by reconstructions of the muon production height by means
of triangulation \cite{Buettner03}.

Another goal of KASCADE-Grande is the development and test of
techniques for radio detection of EAS. Attempts to observe radio
pulses from air showers were made during the late 1960s and they
appear to experience a renaissance at present.  The motivation
arises from the facts that radio signals probe EAS in the maximum
of their longitudinal development, similarly to atmospheric
Cherenkov and fluorescence experiments.  However, radio antennas
would operate and observe EAS 24 hrs a day while optical
observations typically reach a duty cycle of 10\,\% only.
Furthermore, the signals may be cheap to detect with new modern
electronics allowing instrumentation of huge experiments.  LOFAR
(Low Frequency Array) is a new major digital radio interferometer
under development by radio astronomers \cite{lofar}.  Due to its
fully digital nature it will be able to filter out interference
and form beams even after a transient event like an EAS has been
detected.  To test this new technology and demonstrate its
ability to measure air showers, 10 prototype antennas for the
frequency range of 40 to 80 MHz were installed in a joint venture
(LOPES Collaboration) at the site of KASCADE-Grande.
Coincidences observed will allow to reconstruct EAS in a hybrid
mode by both techniques, thereby allowing to judge the
reliability and quality of the radio technique in a unique way
\cite{Horneffer03}.  If proven successful, the radio technique
offers new opportunities for instrumenting the next generation of
giant EAS experiments.

\section{Summary and Outlook}

KASCADE has provided a wealth of new high quality EAS data in the
knee region giving important insight into the origin of the knee
and of CRs in general.  Conclusive evidence has been reached on
the knee being caused by light primaries mostly.  Furthermore,
the data are in agreement with a rigidity scaling of the knee
position giving support to an astrophysical origin by either
maximum acceleration or diffusion/drift models of propagation.
For example, the astrophysical parameters of the GRB model of
Ref.\,\cite{Wick04} are nicely constrained by the preliminary
KASCADE data and are constrained furthermore by measurements of
global anisotropies of CRs.  Observations of a CR excess from the
Monogem SNR cannot be confirmed.

Presently, more data and more observables are being analyzed,
particularly in terms of composition analyses employing
reconstructions of the muon production height.  Together with
measurements of energetic hadrons in the central calorimeter, the
unfolding technique of electron and muon numbers in EAS has
become a powerful tool to reconstruct the properties of primary
particles in EAS and it also provides important clues on how to
improve the hadronic interaction models employed in CORSIKA
simulations.

KASCADE-Grande has just started its routinely data taking and
will extend the measurements up to $10^{18}$\,eV, thereby
allowing to verify the existence of the putative Iron knee
marking the so-called second knee in the all-particle CR
spectrum.  This, together with improved statistics for anisotropy
measurements will allow to confront astro- and particle-physics
motivated models of the knee in much more detail to the
experimental data as is possible now.

The use of radio antennas complementing the experimental
KASCADE-Grande set-up may open a new window to future EAS
observations on large scales.  Interesting first observations
have been made and are presently being analyzed in detail.

\vspace*{5mm} {\small {\bf Acknowledgement:} It is a pleasure to
thank the organizers of the Epiphany meeting for their invitation
and for setting up such an interesting and fruitful meeting in a
very pleasant atmosphere and beautiful hibernal surrounding.
This work has been supported in part by the German Ministry for
Research and Education.}

\end{document}